\documentclass[12pt,preprint]{aastex}

\newcommand{\msun}{M$_\sun$}

\newcommand{\om}{$\Omega_m$}

\newcommand{\ngal}{$N_{gal}$}

\slugcomment{Last modified \today}

\begin{document}

\title{The Richness-Dependent Cluster Correlation Function: \\
Early SDSS Data}

\author{
Neta A. Bahcall\altaffilmark{1},
Feng Dong\altaffilmark{1},
Lei Hao\altaffilmark{1},
Paul Bode\altaffilmark{1},
Jim Annis\altaffilmark{2},
James E. Gunn\altaffilmark{1},
Donald P. Schneider\altaffilmark{3}
}

\altaffiltext{1}{Princeton University Observatory, Princeton, NJ 08544}
\altaffiltext{2}{Fermi National Accelerator Laboratory, P.O. Box 500, Batavia, IL 60510}
\altaffiltext{3}{Dept. of Astronomy and Astrophysics, The Pennsylvania State University, University 
Park, PA 16802}

\begin{abstract}

The cluster correlation function and its richness dependence are determined 
from 1108 clusters of galaxies -- the largest sample of clusters studied 
so far -- found in 379 deg$^2$ of Sloan Digital Sky Survey early data. The results 
are compared with previous samples of optically and X-ray selected clusters. 
The richness-dependent correlation function increases monotonically 
from an average correlation scale of $\sim$ 12 $h^{-1}$ Mpc for poor clusters 
to $\sim$25 $h^{-1}$ Mpc for the richer, more massive clusters with 
a mean separation of $\sim$90 $h^{-1}$ Mpc. X-ray selected clusters 
suggest slightly stronger correlations than optically selected clusters ($\sim$ 2-$\sigma$). 
The results are compared with large-scale 
cosmological simulations. The observed richness-dependent cluster correlation 
function is well represented by the standard flat LCDM model (\om $\simeq$0.3, $h$
$\simeq$0.7), and is inconsistent with the considerably weaker correlations 
predicted by \om = 1 models. An analytic relation for the correlation 
scale versus cluster mean separation, $r_0 - d$, that best describes the observations and the LCDM 
prediction is $r_0 \simeq 2.6 \sqrt{d}$ (for $d \simeq$ 20 - 90 $h^{-1}$ Mpc). Data from 
the complete Sloan Digital Sky Survey, when available, will greatly enhance the 
accuracy of the results and allow a more precise determination of cosmological 
parameters.

\end{abstract}


\section{Introduction} \label{introduction}

The spatial correlation function of clusters of galaxies and its richness dependence 
provide powerful tests of cosmological models: both the amplitude of the correlation 
function and its dependence on cluster mass/richness are determined by the underlying 
cosmology. It has long been shown that clusters are more strongly correlated in space 
than galaxies, by an order-of-magnitude: the typical galaxy correlation 
scale, $\sim$ 5 $h^{-1}$ Mpc, increases to $\sim$ 20 - 25 $h^{-1}$ Mpc for the 
richest clusters (\citealt{bah83, kly83}; see also \citealt{bah88, huc90, pos92, bah92, pea92, 
dal94, cro97, aba98, lee99, bor99, col00, gon02}; and references therein). 
\citet{bah83} showed that the cluster correlation function is richness dependent: the 
correlation strength increases with cluster richness, or mass. Many observations have since confirmed 
these results (references above), and theory has nicely explained them \citep{kai84, 
bah92a, moh96, gov99, colb00, mos00, she01}. But the uncertainties 
in the observed cluster correlation function as manifested by the scatter among different measurements remained large.

In this paper we use the largest sample of clusters yet investigated, 1108 clusters selected 
from 379 deg$^2$ of early Sloan Digital Sky Survey data (see the SDSS cluster catalog: \citealt{bah03}), 
to determine the cluster correlation function. This large, complete sample of objectively selected clusters, 
ranging from poor to moderately rich systems in the redshift range $z$ = 0.1 - 0.3, 
allows a new determination of the cluster correlation function and its richness dependence. 
We compare the SDSS cluster correlation function with results of 
previous optically and X-ray selected clusters (\S \ref{correlation}). We use large-scale cosmological 
simulations to compare the observational results with cosmological models (\S \ref{comparison}). The data are 
consistent with predictions from the standard flat LCDM model (\om $\sim$0.3, $h\sim$0.7), 
which best fits numerous other observations (e.g., \citealt{bah99, ben03, spe03}).

\section{SDSS Cluster Selection} \label{clusterselection}

The Sloan Digital Sky Survey (SDSS; \citealt{yor00}) will provide a comprehensive digital 
imaging survey of 10$^4$ deg$^2$ of the North Galactic Cap (and a smaller, deeper area in 
the South) in five bands ($u$, $g$, $r$, $i$, $z$) to a limiting magnitude of $r<$23, followed 
by a spectroscopic multi-fiber survey of the brightest one million galaxies, to $r<$17.7, 
with a median redshift of $z \sim$0.1 \citep{fuk96, gun98, lup01, hog01, str02}. For more 
details of the SDSS see \citet{smi02}; \citet{sto02}; 
and \citet{pie03}. 

Cluster selection was performed on 379 deg$^2$ of SDSS commissioning data, covering the area 
$\alpha$(2000) = 355\degr\ to 56\degr\ and 145.3\degr\ to 236.0\degr\ at $\delta$(2000)= -1.25\degr\ to 1.25\degr\ 
(runs 94/125 and 752/756). The clusters studied here were selected from these imaging 
data using a color-magnitude maximum-likelihood Brightest Cluster Galaxy method (maxBCG; 
\citealt{ann03}). The clusters are described in the SDSS cluster catalog of \citet{bah03}.
The maxBCG method selects clusters based on the well-known color-luminosity relation of 
the brightest cluster galaxy (BCG) and the E/S0 red ridgeline. 
The method provides a cluster richness estimate, \ngal\ (the number of E/S0 galaxies 
within 1 $h^{-1}$ Mpc of the BCG that are fainter than the BCG and brighter than 
$M_i$(lim) = -20.25), and a cluster redshift estimate that maximizes the cluster likelihood (with 1-$\sigma$ 
uncertainty of $\sigma_z$ = 0.014 for \ngal $\geq$10 and 
$\sigma_z$ = 0.01 for \ngal $\geq$20 clusters). We use all maxBCG clusters in the 
estimated redshift range $z_{est}$ = 0.1 - 0.3 that are above a 
richness threshold of \ngal $\geq$ 10 (corresponding to velocity dispersion $\ga$350 km s$^{-1}$); 
the sample contains 1108 clusters. The selection function and false-positive detection rate 
for these clusters have been estimated from simulations and from visual 
inspection to be $\la10\%$ \citep{bah03}. 

\section{The Cluster Correlation Function} \label{correlation}

The two-point spatial correlation function is determined by comparing the 
observed distribution of cluster pairs as a function of pair separation with the distribution 
of random catalogs in the same volume. The correlation function is estimated from the 
relation $\xi_{cc}(r) = F_{DD}(r)/F_{RR}(r) - 1$, where $F_{DD}$($r$) and $F_{RR}$($r$) are the 
frequencies of cluster-cluster pairs as a 
function of pair separation $r$ in the data and in random catalogs, 
respectively. The random catalogs contain $\sim 10^3$ times the number of clusters 
in each data sample; the clusters are 
randomly positioned on the sky within the surveyed area. 
The redshifts of the random clusters follow the redshifts 
of the observed clusters in order to minimize possible selection effects with redshift. 
Comoving coordinates in a flat LCDM 
cosmology with \om = 0.3 and a Hubble constant of $H_0$ = 100 $h$ km s$^{-1}$ 
Mpc$^{-1}$ are used throughtout.

The uncertainty in the estimated cluster redshifts ($\sigma_z$ = 0.01 for 
\ngal $\geq$20 clusters and $\sigma_z$ = 0.014 for \ngal $\geq$10 to $\geq$15 clusters; 
\S \ref{clusterselection}) causes a small smearing effect in the cluster correlations. 
We use Monte Carlo simulations to correct for this effect. 
We use simulations with a realistic cluster distribution with redshift and richness, 
convolve the clusters with the observed Gaussian scatter in redshift as given above, 
and determine the new convolved cluster correlation function. As expected, the redshift 
uncertainty causes a slight weakening of the true correlation 
function, especially at small separations, due to the smearing effect of the redshift 
uncertainty. We determine the correction factor for this effect as a function of scale 
$r$ from 10$^2$ Monte Carlo simulations for each sample. The correction factor (typically 
$\la 20\%$) is then applied to the correlation function. 
An additional small correction factor due to false-positive detections is also determined 
from Monte Carlo simulations using the estimated false-positive detection rate of 
$10\% \pm5\%$ for \ngal $\geq$10 clusters, $5\% \pm5\%$ for \ngal $\geq$13, and $< 5\%$ for 
the richest clusters with \ngal $\geq$15. The correlation function uncertainties are determined 
from the Monte Carlo simulations. Each simulation contains the same number of clusters 
as the relevant data sample. The final uncertainties include the statistical uncertainties 
and the uncertainties due to the small correction factors in the 
redshift and false-positive corrections.

The correlation function is determined for clusters with richness thresholds of \ngal 
$\geq$10, $\geq$13, $\geq$15, and $\geq$20. The space densities of these clusters, 
corrected for selection function and redshift uncertainty \citep{bah03a}, are 
$5.3\times10^{-5}$, $2.2\times10^{-5}$, $1.4\times10^{-5}$, and $0.5\times10^{-5} \ h^3$ 
Mpc$^{-3}$ ($z_{est}$ = 0.1 - 0.3). The correlation function of the four 
samples are presented in Figure 1 and Table 1. The best-fit 
power-law relation, $\xi$($r$) = ($\frac{r}{r_{0}}$)$^{-\gamma}$, derived for $r$ $\la$ 50 $h^{-1}$ Mpc, 
is shown for each sample. The power-law slope $\gamma$ has been treated both as a free parameter and as a fixed 
value ($\gamma$ = 2). The difference in the correlation scale $r_0$ 
for these different slopes is small ($\la 2\%$), well within the measured uncertainty. 

The richness dependence of the cluster correlation function is shown in Figure 2; it is 
presented as the dependence of the correlation scale $r_0$ on the cluster mean separation $d$ 
\citep{bah83, sza85, bah88, cro97, gov99, colb00}. 
Samples with intrinsically larger mean separations correspond to lower intrinsic cluster 
abundances ($n_{cl}$ = $d^{-3}$) and thus to higher cluster richness and mass (for complete samples).
We compare our results with those of previous optically 
and X-ray selected cluster samples (Figure 2). These include the correlation function of Abell clusters (\citealt{bah83, 
pea92}; Richness class $\geq$1; Richness = 0 clusters are an incomplete sample and should 
not be included); APM clusters \citep{cro97}; Edinburgh-Durham clusters (EDCC; 
\citealt{nic92}); Las-Campanas Distant Cluster Survey (LCDCS; \citealt{gon02}); galaxy groups (2dF; 
\citealt{zan03}); and X-ray selected clusters (REFLEX: \citealt{col00}; and XBACS: 
\citealt{aba98, lee99}). A summary of the results is presented in Table 1.
For proper comparison of different samples, we will use the same set of standard parameters in the 
relative $r_0$ - $d$ plot: redshift $z \sim$0, correlation 
power-law slope $\gamma = 2$, and all scales are in comoving units in the LCDM cosmology.
We discuss these below. 

Most of the cluster samples are at small redshifts, $z \la$ 0.1 (Table 1). The only exceptions 
are the SDSS clusters ($z \simeq$ 0.1 - 0.3), and the LCDCS ($z \simeq$ 0.35 
- 0.575). To convert the results to $z \sim$ 0 we use large-scale cosmological simulations of 
an LCDM model and determine the cluster correlation 
function and the $r_0$ - $d$ relation at different redshifts. 
Details of the simulations and cluster selection are given in \citet{bod01} (see also \S \ref{comparison}). 
The correlation function is determined following the same 
method used for the data. We find that while both $r_0$ and $d$ increase with 
redshift for the same mass clusters, as expected, there 
is no significant change ($\la 3\%$) in the $r_0$ - $d$ relation as the redshift changes from $z$ = 0 to $\sim$ 
0.5 (for $d \sim 20 - 90 \ h^{-1}$ Mpc). In Figure 2 
we plot the individual parameters $r_0$ and $d$ at the sample's 
measured redshift as listed in Table 1; the relative $r_0$ - $d$ 
relation remains essentially unchanged to $z\simeq$ 0. 

Most of the cluster correlation functions (Table 1) have a power-law 
slope in the range of $\sim2\pm0.2$. The APM highest richness 
subsamples report steeper slopes (3.2, 2.8, 2.3); they also have the 
smallest number of clusters (17, 29, 58). The correlation scale $r_0$ is inversely correlated 
with the power-law slope; a steeper slope typically yields a smaller correlation scale. 
We use the APM best $\chi^2$ fit for $r_0$ at $\gamma$ = 2 \citep{cro97} for these 
richest subsamples. Using cosmological simulations, we investigate the dependence of $r_0$ 
on the slope within the more typical observed range of $2\pm0.2$. For the current range of mean 
separations $d$ we find only a small change in $r_0$ ($\la 5\%$) when the slope changes within 
this observed range. The X-ray cluster sample XBACS yields similar correlation scales for slopes 
ranging from $\sim$ 1.8 to 2.5 (\citealt{aba98} and \citealt{lee99}). 
Similarly, the SDSS correlation scales are essentially the same when using a free slope fit (typically 
1.7 to 2.1) or a fixed slope of 2. Since most of the observations are reported with a slope 
of 2, we adopt the latter as the standard slope for the results presented in Figure 2. The only 
correction applied is to the three highest richness APM subsamples; 
these are shown both with and without the correction. We also varify using cosmological 
simulations that the LCDCS sample at $z \sim$ 0.35 - 0.575, with a slope of 2.15, has an $r_0$ - $d$ 
relation consistent with the standard set of parameters used in Figure 2 
($z \simeq$ 0, $\gamma \simeq$ 2). 

Finally, we convert all scales ($r_0$ and $d$ from Table 1) to the same \om =0.3 cosmology (LCDM). 
The effect of the cosmology on the observed $r_0$ - $d$ relation is small ($\la 3\%$), 
partly due to the small redshifts, where the effect 
is small, but also because the cosmology affects both $r_0$ and $d$ in the same way, thus minimizing the 
relative change in the $r_0$ - $d$ relation. 

A comparison of all the results, including the minor corrections discussed above, is shown 
in Figure 2. Figure 2a presents both optically and X-ray selected clusters; 
Figure 2b includes only the optical samples. The richness-dependence of the cluster 
correlation function is apparent in Figure 2. The X-ray clusters 
suggest somewhat stronger correlations than the optical clusters, at a $\sim$2-$\sigma$ level. 
Improved optical and X-ray samples should reduce the scatter 
and help address this important comparison. 

\section{Comparison with Cosmological Simulations} \label{comparison}

We compare the results with large-scale cosmological simulations of a standard LCDM model 
(\om = 0.3, $h$ = 0.67, $\sigma_8$ = 0.9), and a tilted SCDM model, TSCDM (\om = 1, $h$ = 0.5, 
$n$ = 0.625, $\sigma_8$ = 0.5). The TPM high-resolution large-volume simulations \citep{bod01}
used $1.34\times10^8$ particles with an individual particle 
mass of $6.2\times10^{11} h^{-1}$ \msun; the periodic box size is 1000 $h^{-1}$ Mpc for LCDM and 
669 $h^{-1}$ Mpc for TSCDM. The simulated clusters are ordered by their abundance based on cluster mass 
within 1.5 $h^{-1}$ Mpc. The results of the cosmological simulations for the 
$r_0$ - $d$ relation of $z$ = 0 clusters are presented by the two bands in Figure 2 (1-$\sigma$ range). 
A correlation function slope of 2 was used in the analysis. 
We also show the simulations results of \citet{colb00} for 
LCDM, and \citet{gov99} for a standard untilted SCDM (\om = 1, $h$ = 0.5, $n$ = 1, 
$\sigma_8$ = 0.7). The agreement among the simulations is excellent. As expected, 
the untilted SCDM model yields smaller $r_0$'s than the strongly tilted model; 
LCDM yields the strongest correlations.

We determine an analytic relation that approximates the observed and the LCDM $r_0$ - $d$ relation: 
$r_0 \simeq 2.6 \sqrt{d}$ (for $20 \la d \la 90$; all scales are in $h^{-1}$ Mpc). 
The observed richness-dependent cluster correlation function agrees well with the standard LCDM model. 
The correlation scales, and the $r_0$ 
- $d$ relation, increase as \om$h$ decreases and the spectrum shifts to larger scales. The \om = 1 
models yield considerably weaker correlations than observed. 
This fact has of course been demonstrated earlier; in fact, the strength of the cluster correlation 
function and its richness dependence were among the first indications against the standard \om = 1 
SCDM model (\citealt{bah92, cro97, bor99, gov99, colb00}; and references therein).

The scatter in the observed $r_0$ - $d$ relation among different samples is still large, especially 
when both the optical and X-ray selected samples are included. A high-precision 
determination of the cosmological parameters cannot therefore be achieved at this point.

\section{Conclusions}

We determine the cluster correlation function and its richness dependence using 1108 
clusters of galaxies found in 379 deg$^2$ of early SDSS data. 
The cluster correlation function shows a clear richness dependence, with increasing correlation 
strength with cluster richness/mass. The results are combined with previous samples of 
optical and X-ray clusters, and compared with cosmological simulations. We find 
that the richness-dependent cluster correlation function is consistent with predictions from 
the standard flat LCDM model (\om = 0.3, $h$ = 0.7), and, as expected, inconsistent with the 
weaker correlations predicted by \om = 1 models. 
We derive an analytic relation for the correlation scale versus cluster mean separation relation 
that best describes the observations and the LCDM expectations: $r_0 \simeq 2.6 \sqrt{d}$.
X-ray selected clusters suggest somewhat stronger correlations than the optically selected
clusters, at a $\sim$2-$\sigma$ level.

Funding for the creation and distribution of the SDSS Archive has been provided by the 
Alfred P. Sloan Foundation, the Participating Institutions, NASA, the NSF, the U.S. Department 
of Energy, the Japanese Monbukagakusho, and the Max Planck Society. The SDSS Web site is http://www.sdss.org/. 
The SDSS is managed by the Astrophysical Research Consortium (ARC) for the Participating 
Institutions. The Participating Institutions are The University of Chicago, Fermilab, the 
Institute for Advanced Study, the Japan Participation Group, The Johns Hopkins University, Los 
Alamos National Laboratory, the Max-Planck-Institute for Astronomy (MPIA), the Max-Planck-Institute 
for Astrophysics (MPA), New Mexico State University, University of Pittsburgh, Princeton University, 
the United States Naval Observatory, and the University of Washington.

\clearpage


\epsscale{0.92}
\begin{figure}
\plotone{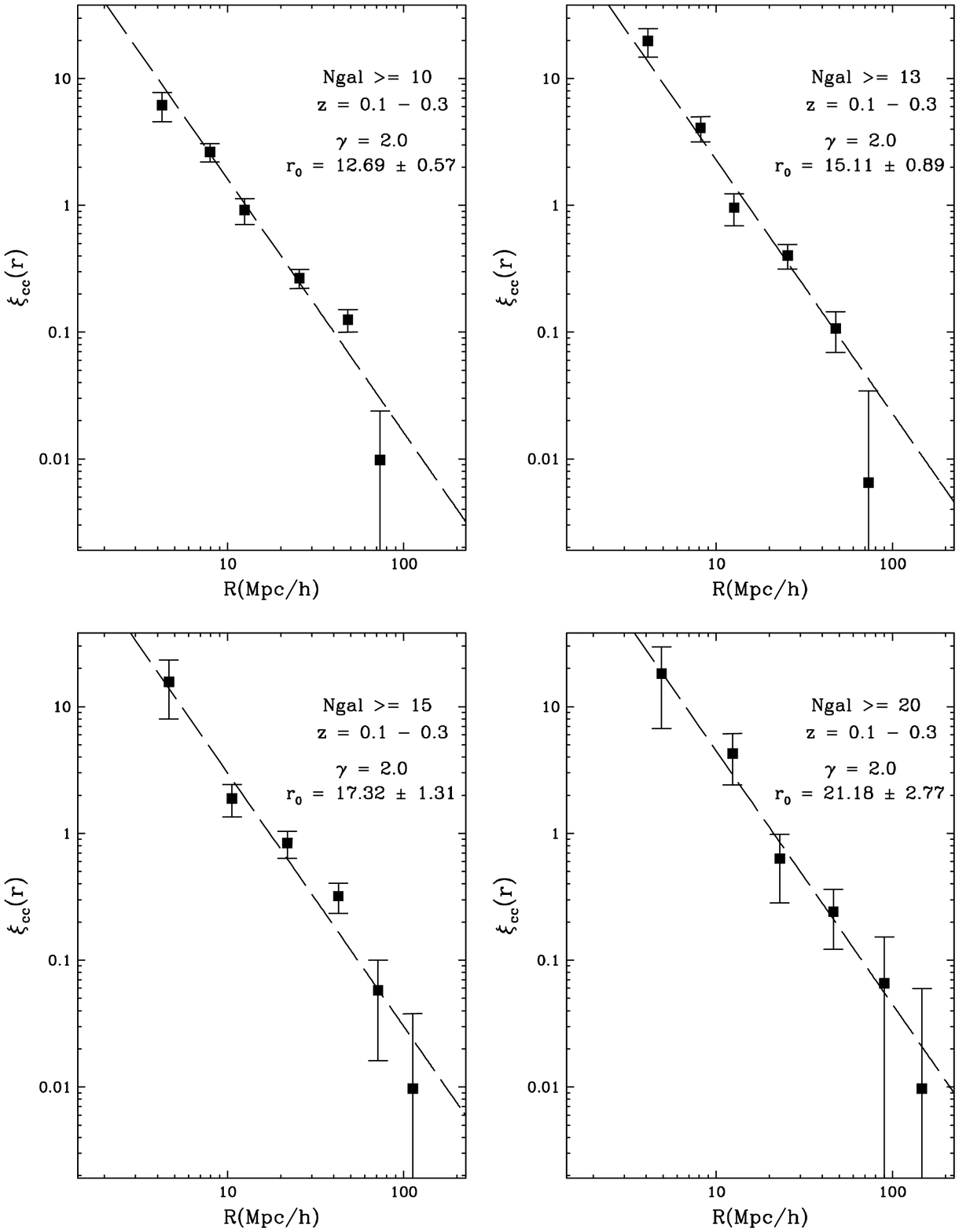}
\caption{The SDSS cluster correlation function for four richness thresholds 
(\ngal $\geq$10, $\geq$13, $\geq$15, $\geq$20). Best-fit functions with slope 2 and 
correlation-scale $r_0$ are shown by 
the dashed lines (1-$\sigma$ uncertainties).
\label{f1}}
\end{figure}

\epsscale{0.92}
\begin{figure}
\plotone{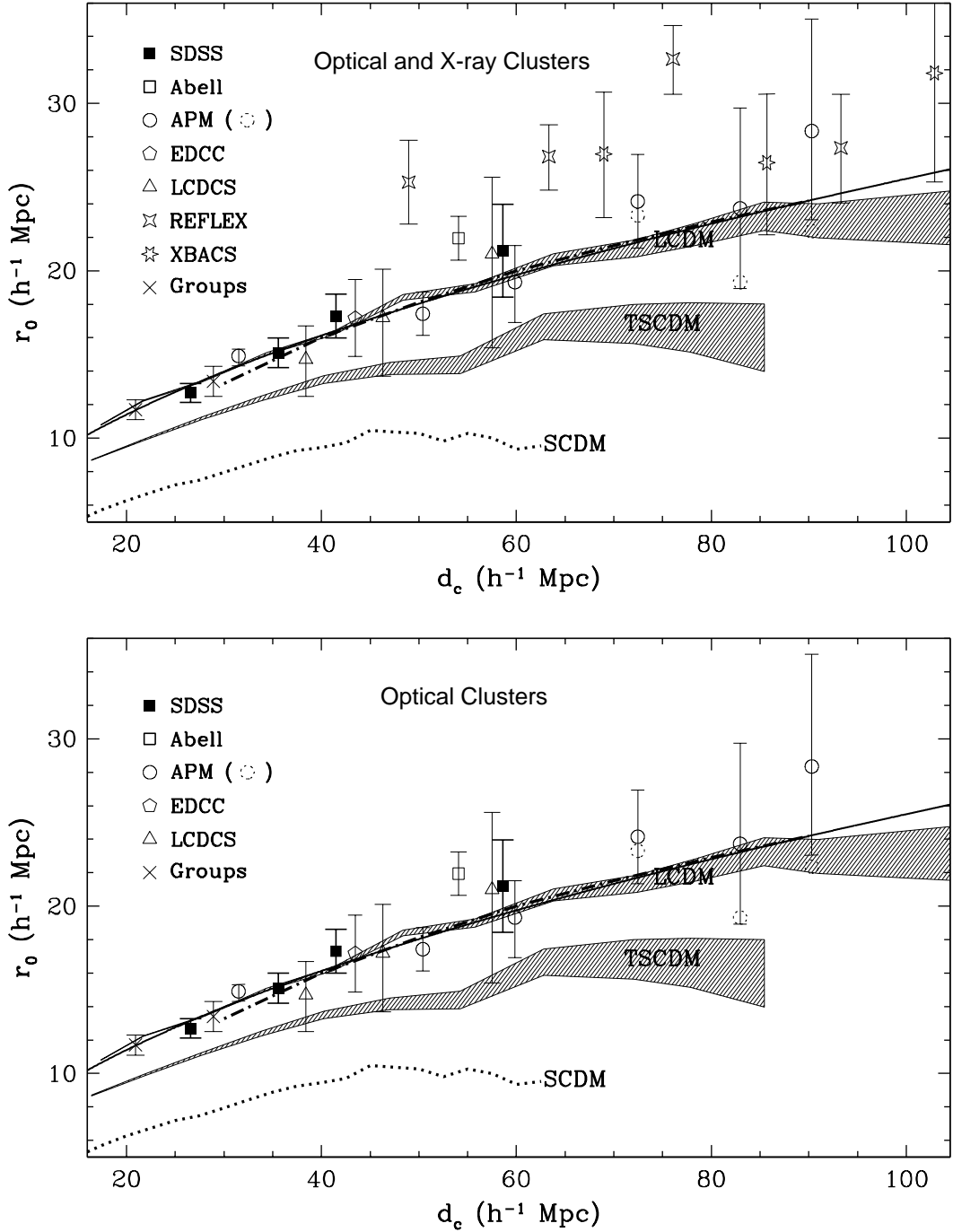}
\caption{Correlation-scale $r_0$ versus mean 
cluster separation $d$ for all samples (Fig.2a) and for optical samples (Fig.2b) (1-$\sigma$ 
uncertainties). A slope $\gamma$= 2 and LCDM comoving 
scales are used (due to conversion to the standard LCDM cosmology, some values differ slightly from 
Table 1; see \S \ref{correlation}). Cosmological simulations 
are presented by the two bands (LCDM and Tilted-SCDM). Previous 
simulations of LCDM (dot-dash) and untilted SCDM (dotted 
curve) are also shown (\S \ref{comparison}). The solid curve represents 
$r_0 = 2.6 \sqrt{d}$ (\S \ref{comparison}).
\label{f2}}
\end{figure}

\clearpage

\begin{deluxetable}{rrrrrrrrr} 
\tabletypesize{\scriptsize}
\tablecolumns{6} 
\tablewidth{0pc} 
\tablecaption{The Cluster Correlation Function} 
\tablehead{ 
\colhead{Sample\tablenotemark{a}} & \colhead{\ $N_{cl}$}  & 
\colhead{$z$} & \colhead{$\gamma$} & \colhead{$r_0$} &
\colhead{$d$} \\
\colhead{} & \colhead{}  & 
\colhead{} & \colhead{} & \colhead{(Mpc)} &
\colhead{(Mpc)}
}
\startdata 	
SDSS \ \ \ \ \ \ \ \ &  &  &  &  &  \\
\ \ N$_{gal}\geq$10 & 1108 & 0.1-0.3 \ & 2 \ \ & $12.7\pm0.6$ & 26.6 \\ 
\ \ N$_{gal}\geq$13 &  472 & 0.1-0.3 \ & 2 \ \ & $15.1\pm0.9$ & 35.6 \\ 
\ \ N$_{gal}\geq$15 &  300 & 0.1-0.3 \ & 2 \ \ & $17.3\pm1.3$ & 41.5 \\ 
\ \ N$_{gal}\geq$20 &  110 & 0.1-0.3 \ & 2 \ \ & $21.2\pm2.8$ & 58.1 \\ 
\hline \\
Abell (1,2) \ \  &  &  &  &  &  \\ 
\ \ Rich $\geq$1 \ & 195 & $\la$0.08 \ \ & 2 \ \ & $21.1\pm1.3$ & 52 \ \\ 
\hline \\
APM (3) \ \ \ \ &  &  &  &  &  \\ 
\ \ R$\geq$50 \ \ & 364 & $\sim<$0.1$>$ & 2.1 \ & $14.2 \ \pm^{0.4}_{0.6}$ & 30 \ \\ 
\ \ R$\geq$70 \ \ & 114 & $<$0.1$>$ \ & 2.1 \ & $16.6\pm1.3$ & 48 \ \\ 
\ \ R$\geq$80 \ \ & 110 & $<$0.1$>$ \ & 1.7 \ & $18.4 \ \pm^{2.2}_{2.4}$ & 57 \ \\ 
\ \ R$\geq$90 \ \ & 58 & $<$0.1$>$ \ & 2.3 \ & $22.2\pm2.8$ & 69 \ \\ 
           &    &           & $[$ 2 $]$ \ & $[ 23.0\pm2.9 ]$\tablenotemark{b} & \\ 
\ \ R$\geq$100 \ & 29 & $<$0.1$>$ \ & 2.8 \ & $18.4\pm4.8$ & 79 \ \\ 
           &    &           & $[$ 2 $]$ \ & $[22.6\pm6.0 ]$\tablenotemark{b} & \\ 
\ \ R$\geq$110 \ & 17 & $<$0.1$>$ \ & 3.2 \ & $21.3\pm5.3$ & 86 \ \\ 
           &    &           & $[$ 2 $]$ \ & $[ 27.0\pm6.7 ]$\tablenotemark{b} & \\ 
\hline \\
EDCC (4) \ \ & 79 & $\la$0.13 \ \ & 2 \ \ \ & $16.2\pm2.3$ & 41 \ \\ 
\hline \\
LCDCS (5) \ & 178 & 0.35-0.475 & 2.15 \ & $14.7 \ \pm^{2.0}_{2.2}$  & 38.4 \\ 
           & 158 & 0.35-0.525 & 2.15 \ & $17.2 \ \pm^{2.9}_{3.5}$  & 46.3 \\ 
           & 115 & 0.35-0.575 & 2.15 \ & $20.9 \ \pm^{4.6}_{5.6}$  & 58.1 \\ 
\hline \\
\ REFLEX (6) &  &  &  &  &  \\  
\ \ L$_{X}\geq$0.08 & 39 & $\la$0.05 \ \ & 2 \ \ & $24.8\pm2.5$ & 48 \ \\ 
\ \ L$_{X}\geq$0.18 & 84 & $\la$0.075 \ & 2 \ \ & $25.8 \ \pm^{1.9}_{2.0}$  & 61 \ \\ 
L$_{X}\geq$0.3 \ & 108 & $\la$0.10 \ \ & 2 \ \ & $31.3 \ \pm^{2.0}_{2.1}$ & 72 \ \\ 
L$_{X}\geq$0.5 \ & 101 & $\la$0.125 \ & 2 \ \ & $25.8 \ \pm^{3.2}_{3.3}$ & 88 \ \\ 
\hline \\
\ XBACS (7,8) &  &  &  &  &  \\  
\ \ L$_{X}\geq$0.24 & 49 & $\la$0.07 \ \ & 1.8-2.5  & $25.7\pm3.7$ & 66 \ \\ 
\ \ L$_{X}\geq$0.48 & 67 & $\la$0.09 \ \ & 1.8-2.5  & $25.2\pm4.1$ & 82 \ \\ 
\ \ L$_{X}\geq$0.65 & 59 & $\la$0.11 \ \ & 1.6-2.2  & $30.3 \ \pm^{8.2}_{6.5}$ & 98 \ \\
\hline \\
Groups (9) \ \ &  &  &  &  &  \\  
\ \ M$_{v}\geq$5e13 & 920 & $<$0.12$>$ & 2 \ \ & $11.7\pm0.6$ & 20.9 \\ 
\ \ M$_{v}\geq$1e14 & 540 & $<$0.13$>$ & 2 \ \ & $13.4\pm0.9$ & 28.9 \\ 
\hline 
\enddata 
\tablenotetext{a}{Sample (with reference), and subsample threshold in richness, X-ray luminosity 
(10$^{44}$ erg s$^{-1}$), or M$_{vir}$ (\msun). References: 1.\citealt{bah83}; 
2.\citealt{pea92}; 3.\citealt{cro97};
[larger $r_0$'s are obtained for APM by \citealt{lee99}];
4.\citealt{nic92}; 5.\citealt{gon02}; 6.\citealt{col00};
7.\citealt{lee99}; see also 8.\citealt{aba98}; 9.\citealt{zan03}. 
The SDSS, LCDCS, and Groups use LCDM cosmology 
for their $r_0$ and $d$; all others use \om=1. All scales are for $h$ = 1.}
\tablenotetext{b}{Correlation-scale r$_0$ using a slope of 2 (see \S \ref{correlation})}.

\end{deluxetable}


\end{document}